\documentclass[12pt]{article}
\usepackage{amsfonts,amsbsy}
\usepackage{amssymb}
\usepackage{graphicx}

\newcommand{\fl}{\hspace*{-5mm}}

\newcommand{\be}{\begin{equation}}
\newcommand{\ee}{\end{equation}}
\newcommand{\ba}{\begin{array}}
\newcommand{\ea}{\end{array}}
\newcommand{\p}{\partial}
\newtheorem{theorem}{Theorem}
\DeclareMathAlphabet{\bi}{OML}{cmm}{b}{i}
\begin{document}

\title{\vspace{-1.5cm}{\bf Maximal superintegrability of Benenti systems}}

\author{%
Maciej B\l aszak\dag\ and Artur Sergyeyev\ddag\\[3mm] %
%\address{
\dag\ Institute of Physics, A. Mickiewicz University\\
Umultowska 85, 61-614 Pozna\'{n}, Poland\\[3mm]
\ddag\ Silesian University in Opava, Mathematical Institute,\\ 
Na Rybn\'\i{}\v{c}ku 1, 746\,01 Opava, Czech Republic\\[3mm] 
E-mail: {\tt{blaszakm@amu.edu.pl}} {\protect{\rm and}} {\tt{Artur.Sergyeyev@math.slu.cz}
}}
\date{}
\maketitle

\begin{abstract}
For a class of Hamiltonian systems naturally arising in the modern theory
of separation of variables, we establish their maximal superintegrability by
explicitly constructing the additional integrals of motion.
\end{abstract}

%\jl{1}

%\pacs{02.30Ik, 45.20Jj} \ams{70H06, 70G45, 37J35}

%\submitto{\JPA}

%\maketitle

For a Hamiltonian dynamical system on a $2n$-dimensional phase space to be
completely integrable, the classical Liouville theorem requires $n$
involutive integrals of motion. Solving the equations of motion then amounts
to $n$ quadratures, but performing the latter still can be a challenge.
However, if there are $m>n$ integrals of motion, only $2n-m$ quadratures are
required. In particular, for the so-called \emph{maximally superintegrable}
systems (i.e., those possessing $2n-1$ integrals of motion) we need just
\emph{one} quadrature. 
\looseness=-1

Superintegrable systems on two- and three-dimensional Euclidean spaces and
spaces of constant curvature were extensively studied in the literature, see
e.g.\ \cite{fri65}--\cite{gra02} and also \cite{kaln02} for the case of
nonconstant curvature. On the other hand, there is not much known about
superintegrability in higher dimensions. Remarkable exceptions include $n$%
-dimensional Winternitz--Smorodinsky model \cite{fri65} and its
generalizations \cite{bal03}, Calogero--Moser--Sutherland system \cite%
{rw83,gon98} and the systems with isochronic potentials \cite{gon04} and
modified Coulomb potential \cite{rod02}. Notice that for the majority of
these systems the separation of variables takes place in the original
coordinates and the Hamiltonians themselves are additively separable.
\looseness=-1

In this letter we consider the St\"ackel systems from the class described by
Benenti \cite{ben93, ben97}, so the corresponding Hamiltonians are quadratic
in momenta. We show that some of these systems (namely those associated with
flat or constant curvature metrics) are maximally superintegrable for
arbitrary dimension $n$ of the configuration space, and it seems plausible
that they are multiseparable too.

Consider an $n$-dimensional manifold $M$ with local coordinates $%
q^1,\dots,q^n$ and the $n \times n$ matrices
\[
\fl G=\left(
\begin{array}{cccccc}
0 & 0 & 0 & 0 & 0 & 1 \\
0 & \cdots & 0 & 0 & 1 & q^{1}\\
0 & \cdots & 0 & 1 & q^{1} & q^{2}\\
\vdots & \vdots & \vdots & \vdots & \ddots & \vdots\\
0 & 1 & q^{1} & q^{2}  & \cdots & q^{n-2}\\
1 & q^{1} & q^{2} & q^{3} & \cdots & q^{n-1}%
\end{array}
\right) \quad\mbox{and}\quad 
L=\left(
\begin{array}{cccccc}
-q^{1} & 1 & 0 & 0 & \cdots & 0 \\
-q^{2} & 0 & 1 & 0 & \cdots & 0\\
-q^{3} & 0 & 0 & 1 & \cdots & 0\\
\vdots & \vdots & \vdots & \vdots & \ddots & \vdots\\
-q^{n-1} & 0 & 0 & \cdots & 0 & 1\\
-q^{n} & 0 & 0 & 0 & \cdots & 0%
\end{array}
\right).
\]
If we interpret $G$ as a contravariant metric on $M$, then $L$ is a
special conformal Killing tensor for $G$ in the sense of Crampin and
Sarlet \cite{cra} (and moreover, this holds 
for any metric of the form $L^i G$,
$i\in\mathbb{Z}$ \cite{bl03}), and hence
%Let $(q^1,\dots,q^n,p_1,\dots,p_n)$ be local coordinates on the cotangent bundle $T^*M$.
%It can be shown  \cite{} that
for any integer $i$ the quantities
\begin{equation}
E_{i,r}=\frac{1}{2}\sum_{l,m=1}^{n}(K_{r}L^{i} G)^{lm}p_{l}p_{m},\quad
r=1,...,n,   \label{Blaszak:3.4}
\end{equation}
are in involution with respect to the canonical Poisson bracket in $(p,q)$%
-variables: $\{q^i,p_j\}=\delta^i_j$, $i,j=1,\dots,n$. Here
$(q^1,\dots,q^n,p_1,\dots,p_n)$ are local coordinates on the
co\-tangent bundle $T^*M$ and the Killing tensors $K_{r}$ are
constructed from $L$ as follows~\cite{ben93}: %in $p,q$-coordinates:
\[
K_1=\mathbb{I}, \quad K_{r+1}=\sum_{k=0}^{r}q^{k}L^{r-k}, \quad
r=1,\dots,n-1,
\]
where we set for convenience $q^0\equiv 1$ and $\mathbb{I}$ stands for the $%
n \times n$ unit matrix. \looseness=-1 Notice that $q^{r}$ are coefficients
of the characteristic polynomial of $L$:
\begin{equation}
\det(\xi I-L)=\xi^{n}+q^{1}\xi^{n-1}+...+q^{n}.   \label{Blaszak:2.8b}
\end{equation}
\looseness=-2

The Hamiltonians $E_{i,r}$ belong to the family of St\"ackel separable
systems, and the separation of variables is achieved by passing to the
coordinates $(\mu,\lambda)$ related to $(p,q)$ by %means of
the formulas
\[
p_i=-\sum\limits_{k=1}^n (\lambda^k)^{n-i}\mu_k/\Delta_k, \quad q^i=(-1)^i
\sigma_i(\lambda), \quad i=1,\dots,n,
\]
where $\sigma_i$ is an $i^\mathrm{th}$ symmetric polynomial in the variables 
$\lambda^1,\dots,\lambda^n$ ($%
\sigma_0=1$, $\sigma_1=\sum_{i=1}^n\lambda^i,\dots,
\sigma_n=\lambda^1\lambda^2\cdots\lambda^n$) and $\Delta_i=\prod_{j=1,j\neq
i}^n (\lambda^i-\lambda^j)$.

%In fact, the Hamiltonians $E_{i,r}$ belong to the class studied by Benenti [].
%Notice also that $L$ is a special conformal Killing tensor in the sense of Crampin \cite{}
%for any (contravariant)
%metric $G_i\equiv L^i G$, $i\in\mathbb{Z}$.

%The above formulas for $E_{i,r}$ and $K_r$,
%as well as the proof of commutativity of $E_{i,r}$ for a given $i$
%are due to Benenti who, however, worked with all these objects in
%the separation variables $(\lambda,\mu)$. In particular, $L$ is nothing but
%the so-called special conformal Killing tensor, see \cite{}.

The main advantage of using the `nonstandard' variables $(p,q)$ is that
%the Hamiltonians
$E_{i,r}$ are polynomial in $q$'s for $i\geq 0$, so %in this case
it is natural to search for additional integrals of motion that also are
polynomial in $p$'s and $q$'s.
\looseness=-1

For $i=0,\dots,n$ the (contravariant) metrics $(G_i)^{rs}\equiv (L^i G)^{rs}$
are flat and we have %can write $E_{i,1}$ as follows:
\[
\fl
\begin{array}{l}
E_{i,1}=\frac12\sum\limits_{k=0}^{n-i-1} q^{k} \sum
\limits_{j=k+1}^{n-i}p_{j}p_{n-i+k-j+1}-\frac12\sum\limits_{k=1}^{i}
q^{n-i+k} \sum\limits_{j=1}^{k}p_{n-i+j}p_{n-i+k-j+1}.%
\end{array}
\]
The metric $G_{n+1}$ has nonzero constant curvature and we obtain
\[
\begin{array}{l}
E_{n+1,1}=\frac12\sum\limits_{i,j=1}^{n} (q^{i}q^{j}-q^{i+j})p_i p_j,%
\end{array}
\]
where we assume that $q^{k}\equiv 0$ for $k>n$.

%Let $\{,\}$ stand for the canonical Poisson bracket in $(q,p)$-
%variables: $\{q^i,p_j\}=\delta^i_j$, $i,j=1,\dots,n$.
We readily find that
\begin{eqnarray*}
\{E_{i,1},p_{n-i}\} &=&0,\quad i=0,\dots ,n-1, \\
\{E_{i,1},q^{n}p_{n-i+2}\} &=&0,\quad i=2,\dots ,n+1, \\
\{E_{i,1},q^{1}\} &=&-p_{n-i},\ \ i=0,...,n-1. 
\end{eqnarray*}%
Thus $q^{n-i}$ is a cyclic coordinate for $E_{i,1}$, $i=0,\dots ,n-1$, and $%
F_{i,r}=\{E_{i,r},p_{n-i}\}$ for $r=2,\dots ,n$ provide $(n-1)$ additional,
quadratic in momenta, integrals for the dynamical systems associated with
the Hamiltonians $E_{i,1}$, $i=0,\dots ,n-1$.
%and so do $\tilde F_{i,r}=\{ E_{i,r},q^n p_{n-i+2}\}$
%for $E_{i,1}$ with $i=n,n+1$.
%It can be shown that $\tilde E_{i,r}$ are functionally independent

Consider $s$ functions $f_i$ of the form
\[
f_i=\sum\limits_{j,k=1}^n A_i^{jk}(q)p_j p_k
%+\sum\limits_{j=1}^n B_i^{j}(q)p_j
+C_i(q).
\]
%where $C_i(q)$ are arbitrary functions of $q^1,\dots,q^n$.
It is easily seen that if the matrices $A_i$ are linearly
independent over the field of all locally analytic functions of
$q^1,\dots,q^n,p_1,\dots,p_n$, then
$f_i$ are \emph{functionally independent}. %
%
%To prove this, assume the converse. Then there exists a function
%$M\not\equiv 0$ of $s$ arguments such that $M(f_1,\dots, f_s)\equiv
%0$, and therefore $\p M(f_1,\dots, f_n)/\p p_j=2
%\sum_{m=1}^s\sum_{k=1}^n\p M/\p f_m A_{m}^{jk}(q) p_k \equiv 0$ for
%all $j$.

%A fairly straighforward but rather tedious computation shows that
Using this result we readily find that $(2n-1)$ quantities
$E_{i,s}$, $s=1,\dots,n$,
and $F_{i,r}$, $r=2,\dots,n$ are functionally independent for any $%
i=0,\dots,n-1$, and hence the Hamiltonians $E_{i,1}$ are maximally
superintegrable for all these $i$. % $i=0,\dots,n-1$.
The same is true for $i=n$ and $i=n+1$, but in this case the additional
integrals have the form $F_{i,r}=\{ E_{i,r},q^n p_{n-i+2}\}$, $%
r=2,\dots,n$,
%. Notice that $\tilde F_{i,r}$
and they  again are quadratic in momenta. %
\looseness=-1

This picture can be extended to the case of nonzero potentials. Namely,
consider the basic separable potentials given by the recursion relations
\cite{bl98,bl01}
\begin{equation}
\begin{array}{l}
\fl V_{r}^{(m+1)}=V_{r+1}^{(m)}+V_{r}^{(1)}V_{1}^{(m)},\quad
m=1,2,\dots, \quad
 V_{r}^{(1)}=-q^r, \\
\fl V_r^{(0)}=0, \\
\fl V_{r}^{(-m-1)}=V_{r-1}^{(-m)}+V_{r}^{(-1)}V_{n}^{(-m)}, \quad
m=1,2,\dots,\quad V_{r}^{(-1)}=-q^{r-1}/q^n.\label{recursion}%
\end{array}%
\end{equation}
As the potentials $V_{1}^{(m)}$ are independent of $q^{j}$ for
$m=j-n,\dots ,j-1$,
%$\p V_{1}^{(m)}/\p q^{j}=-1$ for $m=j$ and $\p
%V_{1}^{(m)}/\p q^j=2q^{1}$ for $m=j+1$,
the above analysis can be
generalized to yield the following result:

\begin{theorem}
For any natural $n\geq 2$ the dynamical systems with the Hamiltonians $%
H_{i,1}^{(k)}=E_{i,1}+V_{1}^{(k)}$ are maximally superintegrable for all $%
i=0,\dots ,n-1,$ $k=-i,...,n-1-i$  and $i=n,n+1$, $k=2-i,...,n+1-i$.
%and $i=n,n+1$, $k=-i+2,\dots,n+1-i$ .
Namely, besides $n$ involutive integrals
$H_{i,r}^{(k)}=E_{i,r}+V_{r}^{(k)}$, $r=1,\dots ,n$,
the dynamical system associated with %the Hamiltonian
$H_{i,1}^{(k)}$ has $(n-1)$ additional integrals $F_{i,s}^{(k)}$,
$s=2,\dots,n$, of the following form:
\begin{itemize}
\item[a)]
%commutes with
$F_{i,s}^{(k)}=\{H_{i,s}^{(k)},p_{n-i}\}$, $s=2,\dots ,n$, for
$i=0,\dots,n-1$, $k=-i,...,n-1-i$,
%\item[b)]
%$F_{i,s}^{(n-i)}=\{H_{i,s}^{(n-i)},\frac{1}{2}%
%p_{n-i}^{2}-q^{1}\}$ for $i=0,\dots,n-1$, $k=n-i$,
%\item[c)]
%$F_{i,s}^{(n-i+1)}=\{H_{i,s}^{(n-i+1)},\frac{1}{2}%
%p_{n-i}^{2}+(q^{1})^{2}\}$ for $i=0,\dots,n-1$, $k=n-i+1$,
\item[b)]
$F_{i,s}^{(k)}=%
\{H_{i,s}^{(k)},q^{n}p_{n-i+2}\}$ for $i=n,n+1$, $k=2-i,...,n+1-i$,
\end{itemize}
and these $(2n-1)$ integrals ($H_{i,r}^{(k)}$, $r=1,\dots,n$, and
$F_{i,s}^{(k)}$, $s=2,\dots,n$) are functionally independent.
\end{theorem}

%Notice that the (contravariant) metrics $G_i$
%are flat for $i=0,\dots,n$ and have constant curvature for $i=n+1$.
%, while for $i<0$ and $i>n+1$ they have nonconstant nonzero quadrature.

%\section*{Acknowledgements}

Let us illustrate this theorem by a nontrivial example. First, let $%
n=4$, $i=4$, $k=-2$. The functions
\begin{eqnarray*}
\fl H_{4,1}^{(-2)} =-\frac{1}{2}q^{1}{p_{{1}}}^{2}-q^{2}p_{{1}}p_{{2}%
}-q^{3}p_{{1}}p_{{3}}-q^{4}p_{{1}}p_{{4}}-\frac{1}{2}q^{3}{p_{{2}}}%
^{2}-q^{4}p_{2}p_{{3}}+{q^{3}}/{(q^{4})^{2}}, \\
\fl H_{4,2}^{(-2)} =-\frac{1}{2}q^{2}{p_{{1}}}^{2}-q^{3}p_{{1}}p_{{2}%
}-q^{4}p_{{1}}p_{{3}}+\frac{1}{2}(q^{2})^{2}{p_{{2}}}^{2}+q^{2}q^{3}p_{{2}%
}p_{{3}}+q^{2}q^{4}p_{{2}}p_{{4}}-\frac{1}{2}q^{1}q^{3}{p_{{2}}}^{2} \\
\hspace*{-5mm}-p_{{2}}q^{1}q^{4}p_{{3}}-\frac{1}{2}q^{4}{p_{{2}}}^{2}+\frac{1}{2}%
(q^{3})^{2}{p_{{3}}}^{2}+q^{3}q^{4}p_{{3}}p_{{4}}+\frac{1}{2}(q^{4})^{2}{p_{{%
4}}}^{2}+q^{1}q^{3}/(q^{4})^{2}-1/q^{4}, \\
\fl H_{4,3}^{(-2)} =-\frac{1}{2}q^{3}{p_{{1}}}^{2}-q^{4}p_{{1}}p_{{2}}+%
\frac{1}{2}q^{3}q^{2}{p_{{2}}}^{2}+(q^{3})^{2}p_{{2}}p_{{3}}+q^{3}q^{4}p_{{2}%
}p_{{4}}-\frac{1}{2}q^{1}q^{4}{p_{{2}}}^{2}+q^{4}q^{3}{p_{{3}}}^{2} \\
\hspace*{-5mm}+(q^{4})^{2}p_{{3}}p_{{4}}-q^{1}/q^{4}+q^{2}q^{3}/{(q^{4})^{2}}, \\
\fl H_{4,4}^{(-2)} =-\frac{1}{2}q^{4}{p_{{1}}}^{2}+\frac{1}{2}q^{4}q^{2}{%
p_{{2}}}^{2}+q^{4}q^{3}p_{{2}}p_{{3}}+(q^{4})^{2}p_{{2}}p_{{4}}+\frac{1}{2}%
(q^{4})^{2}{p_{{3}}}^{2}-q^{2}/q^{4}+(q^{3})^{2}/(q^{4})^{2}
\end{eqnarray*}%
are in involution by construction. Moreover, $H_{4,1}^{(-2)}$
commutes with $q^{4}p_{2}$ and
thus by Theorem 1 the quantities %\begin{eqnarray*}
\[
\fl F_{4,2}^{(-2)}=\frac{1}{2}q^{4}p_{1}^{2}, \quad
F_{4,3}^{(-2)}=\frac{1}{2}
q^{4}q^{3}p_{2}^{2}+(q^{4})^{2}p_{2}p_{3}-q^{3}/q^{4},\quad F%
_{4,4}^{(-2)}=\frac{1}{2}(q^{4})^{2}p_{2}^{2}+1,
\]%
are additional integrals of motion. %\end{eqnarray*}
As $H_{4,1}^{(-2)},H_{4,2}^{(-2)},H_{4,3}^{(-2)},H_{4,4}^{(-2)},
F_{4,2}^{(-2)},F_{4,3}^{(-2)},F_{4,4}^{(-2)}$ are
functionally independent, the dynamical system associated with $%
H_{4,1}^{(-2)}$ is maximally superintegrable. Notice that all
$(2n-1)$ integrals in this case are quadratic in momenta.

For all natural $n\geq 2$ we have $\p V_{1}^{(m)}/\p q^{j}=-1$ for
$m=j$ and $\p V_{1}^{(m)}/\p q^j=2q^{1}$ for $m=j+1$, whence
$\{H_{i,1}^{(n-i)},\frac{1}{2} p_{n-i}^{2}-q^{1}\}=0$ and
$\{H_{i,1}^{(n-i+1)},\frac{1}{2} p_{n-i}^{2}+(q^{1})^{2}\}=0$ for
$i=0,\dots,n-1$. Therefore the dynamical systems associated with
$H_{i,1}^{(n-i)}$ and $H_{i,1}^{(n-i+1)}$ for $i=0,\dots,n-1$
possess $(n-1)$ additional integrals,
%$F_{i,s}^{(k)}$, $s=2,\dots,n$
%also for $i=0,\dots,n-1$ and $k=n-i$ and $n-i+1$,
namely $F_{i,s}^{(n-i)}=\{H_{i,s}^{(n-i)},\frac{1}{2}
p_{n-i}^{2}-q^{1}\}$, $s=2,\dots,n$, and
$F_{i,s}^{(n-i+1)}=\{H_{i,s}^{(n-i+1)},\frac{1}{2}
p_{n-i}^{2}+(q^{1})^{2}\}$, $s=2,\dots,n$, respectively. However, in
these cases the additional integrals are {\em cubic} in momenta, so
%,unlike Theorem 1,
we were unable to prove the functional independence of the sets
($H_{i,r}^{(k)}$, $r=1,\dots,n$, and $F_{i,s}^{(k)}$, $s=2,\dots,n$)
for $i=0,\dots,n-1$ and $k=n-i$ and $k=n-i+1$ in full generality so
far.
%using same technique as above.
Nevertheless we are certain that
these sets indeed are functionally independent for all these values
of $i$ and $k$, and hence the dynamical systems associated with
$H_{i,1}^{(n-i)}$ and $H_{i,1}^{(n-i+1)}$ are maximally
superintegrable.

For instance, let $n=3$, $i=0$, $k=4$. The functions
\begin{eqnarray*}
\fl H_{0,1}^{(4)}=p_{3}p_{1}+\frac{1}{2}p_{2}^{2}+q_{1}p_{2}p_{3}+\frac{1}{2}%
p_{3}^{2}q^{2}+2q^{3}q^{1}-3q^{2}(q^{1})^{2}+(q^{2})^{2}+(q_{1})^{4}, \\
\fl H_{0,2}^{(4)}
=p_{1}p_{2}+q_{1}p_{1}p_{3}+q_{1}p_{2}^{2}+(q^{1})^{2}p_{2}p_{3}-\frac{1}{2%
}q^{3}p_{3}^{2}+\frac{1}{2}q^{1}q^{2}p_{3}^{2}-(q^{1})^{2}q^{3}+2q^{2}q^{3}
\\
-2q^{1}(q^{2})^{2}+(q^{1})^{3}q^{2}, \\
\fl H_{0,3}^{(4)} =\frac{1}{2}%
p_{1}^{2}+q^{1}p_{1}p_{2}+q^{2}p_{1}p_{3}-q^{3}p_{2}p_{3}+\frac{1}{2}%
(q^{1})^{2}p_{2}^{2}+q^{1}q^{2}p_{2}p_{3}-\frac{1}{2}q^{1}q^{3}p_{3}^{2} \\
+\frac{1}{2}(q^{2})^{2}p_{3}^{2}+q^{3}(-2q^{2}q^{1}+q^{3}+(q^{1})^{3})
\end{eqnarray*}%
again are in involution by construction; $H_{0,1}^{(4)}$ also commutes with $\frac{1}{2}%
p_{3}^{2}+(q^{1})^{2}$, and hence by the above so do
\begin{eqnarray*}
F_{0,2}^{(4)} &=&-2q^{1}p_{2}-3(q^{1})^{2}p_{3}-\frac{1}{2}%
p_{3}^{3}+2q^{2}p_{3},\  \\
F_{0,3}^{(4)}
&=&-q^{1}p_{1}-2(q^{1})^{2}p_{2}-4q^{1}q^{2}p_{3}-p_{2}p_{3}^{2}-\frac{1}{2}%
q^{1}p_{3}^{2}+2q^{3}p_{3}+(q^{1})^{3}p_{3}.
\end{eqnarray*}%
As
$H_{0,1}^{(4)},H_{0,2}^{(4)},H_{0,3}^{(4)},F_{0,2}^{(4)},F_{0,3}^{(4)}$
are functionally independent, the dynamical system associated with
$H_{0,1}^{(4)}$ is maximally superintegrable as well. In contrast
with the previous example, the additional integrals
$F_{0,2}^{(4)},F_{0,3}^{(4)}$ are {\em cubic} in momenta.

%\ack 
\subsection*{Acknowledgements}
This research was partially supported by the Czech Grant Agency (GA%
\v{C}R) under grant No.\ 201/04/0538, Ministry of Education, Youth
and Sports of the Czech Republic under grant MSM:J10/98:192400002
and the development project No. 254/b for the year 2004, KBN
Research Grant No.\ 1 PO3B 111 27, and by the Silesian University in
Opava through the internal grant IGS 1/2004. MB is pleased to
acknowledge kind hospitality of the Mathematical Institute of
Silesian University in Opava. The authors thank the anonymous
referee for useful suggestions. \looseness=-1

%\section*{References}

\end{document}